\documentclass[aps,prl,twocolumn,showpacs,groupedaddress,preprintnumbers,amsmath,amssymb]{revtex4}

\usepackage {amsmath,amssymb,epsfig,color}
\usepackage {graphicx}
\usepackage {dcolumn}
\usepackage {bm}

\begin{document}

\title{EXAMINATION OF THE ALLOCATIONS OF BUILDING MOLECULES IN THE SINGLE CRYSTAL OF THE PARA-DICHLORBENZOL WITH THE P-DIBROMOBENZENE SOLID SOLUTION BY THE METHOD OF THE RAMAN EFFECT OF LIGHT DEPENDING ON REQUIREMENTS OF SELECTION}

\author {M.A.~Korshunov}
 \email {mkor@iph.krasn.ru}
 \affiliation {L.V. Kirensky Institute of Physics, Siberian Branch of Russian Academy of Sciences, 660036 Krasnoyarsk, Russia}

\date{\today}

\begin{abstract}
Allocation of molecules of para-dichlorbenzol in equimolar single crystals 
of para-dichlorbenzol with p-dibromobenzene solid solutions grown by the 
Bridgmen's method is studied. It is shown, that the mutual concentration of 
builders longwise an exemplar depends on requirements of selection. Probably 
as a uniform modification of concentration of builders along an exemplar, 
and a wavy modification of concentration. Critical speed at a modification 
of character of allocation made 15.0 $\cdot $10-6 cm/s at a lapse rate of 
temperature dT/dl=7.7 grad/cm.
\end{abstract}

\maketitle

In a number of publications [1] it is shown, that for two-component solid 
solutions depending on requirements of selection of single crystals on 
method Bridgmen it is possible both a uniform modification of concentration 
of builders, and wavy. In operation it speaks periodic appearance and 
wedging out of blocks during propagation boundaries which segregation of 
impurity takes place. These examinations were spent on junction$_{ 
}$Ñ$_{14}$H$_{14}$ and$_{ }$C$_{14}$H$_{10}$. Operation by [2] us it is 
carried out examination of solid solutions of para-dichlorbenzol with a 
p-dibromobenzene grown on method Bridgmen at velocities of pulling down of 
dish V=8.3$\cdot $10-6--8.9$\cdot $10-6 cm/s and a lapse rate temperature 
dT/dl=7.6-7.7 grad/cm. Monotone allocation of impurity has been discovered. 
Therefore in the yielded operation are carried out examinations for various 
velocities of pulling down of dish. To define nonmonotone allocation of 
impurity in studied mix-crystals takes place at a modification of 
requirements of selection.

It promotes understanding of the mechanism of crystal growth and will enable 
to influence it. 

P-dibromobenzene and para-dichlorbenzol ($\alpha $-modification), 
crystallizes in centrosymmetric space group P2$_{1}$/a with two 
molecules in a low level cell. These chips were researched by a X-ray 
diffraction method [3], methods of a Raman effect of light [4] and a nuclear 
quadrupole resonance [5]. X-ray diffraction data on the mix-crystals studied 
in operation, us it is not revealed. These mix-crystals have been selected 
because their components are isomorphic among themselves and form solid 
solutions at any concentrations of builders. Single crystals of solid 
solutions have been grown on method Bridgmen.

In the yielded operation the method of a Raman effect of light which allows 
to judge on spectrums of the lattice and intramolecular oscillations 
character of layout of molecules of impurity among molecules of the main 
chip is used. 

\begin{figure}
\includegraphics[width=1\linewidth]{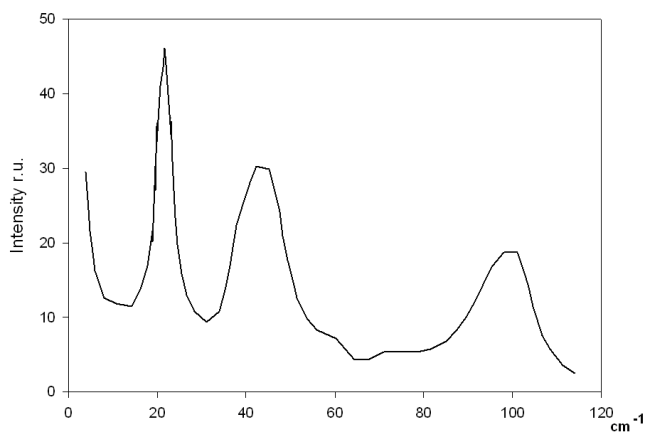}
\caption{Spectrum of the lattice oscillations.}
\label{fig1}
\end {figure}

In Fig.~\ref{fig1} spectrum of the lattice oscillations the mixed chip of a 
p-dibromobenzene with para-dichlorbenzol is reduced at concentration 50 
pier. {\%} of builders. The spectrum of mix-crystals is similar to spectrums 
of builders that speaks about allocation of impurity as substitution. 

The value of concentration of impurity in the grown single crystals of solid 
solutions was defined on relative intensity of the valence intramolecular 
oscillations. According to operation [6] line in spectrum ÊÐÑ of a 
p-dibromobenzene with frequency of $\nu $ =212.0 cm$^{ - 1}$ corresponds to 
valence vibration C-Br, and a line with frequency of $\nu $ =327.0 cm$^{-1}$ 
in para-dichlorbenzol -- to valence vibration C-Cl.

\begin{figure}
\includegraphics[width=1\linewidth]{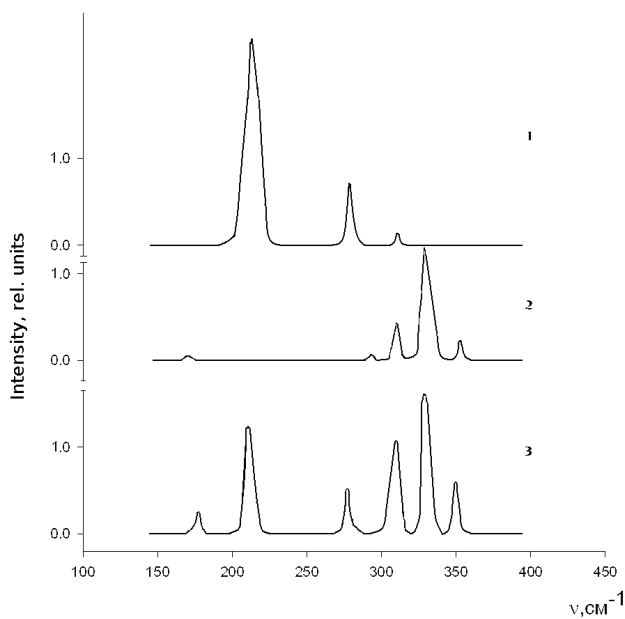}
\caption{Spectrum of intramolecular oscillations.}
\label{fig2}
\end {figure}

In Fig.~\ref{fig2} spectrums of intramolecular oscillations (in the field of from 
150 up to 400 cm$^{ - 1})$ a p-dibromobenzene (1), para-dichlorbenzol (2) 
and studied mix-crystals are presented at equimolecular concentrations of 
para-dichlorbenzol in a p-dibromobenzene (3). Using a relation intensity the 
symmetric valence vibrations in mix-crystals, the modification of 
concentration of builders longwise single crystals is revealed. Examinations 
of different exemplars it was spent at the same parameters of a 
data-acquisition equipment. In operation concentration of builders was 
measured in molar unities.

The single crystal was grown in a glass tube in diameter d = 1 cm and length 
h = 10 cm with the plucked capillary. In a handset put mother substances in 
the necessary percentage. After that from a handset pumped out air, and she 
was sealed off.

The handset with substance was omitted in the crystallizing furnace with 
velocity V=8.3$\cdot $10-6--8.9$\cdot $10-6 cm/s. The lapse rate of 
temperature of the furnace was set by various winding of a heating coil and 
made dT/dl=7.6-7.7 a grad/cm.

\begin{figure}
\includegraphics[width=1\linewidth]{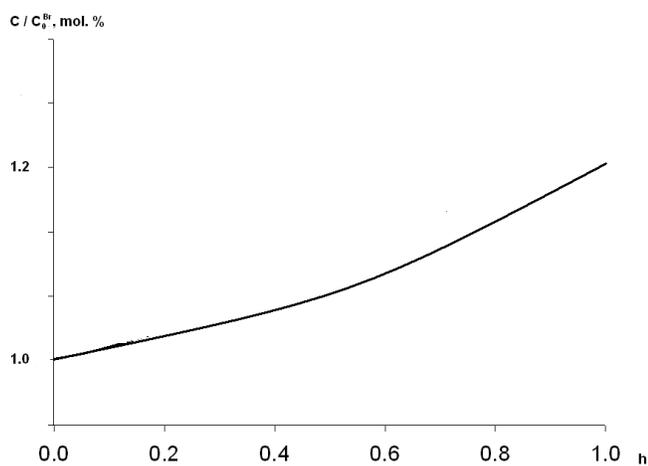}
\caption{Allocation of impurity of para-dichlorbenzol and 
p-BromoClorobenzene longwise (h) the mixed chip at low speed of propagation.}
\label{fig3}
\end {figure}

In Fig.~\ref{fig3} association of relative concentration of a p-dibromobenzene to 
para-dichlorbenzol is shown. The initial concentration of 
p-dibromobenzene C$^{Br}_{0}$ in fusion mixture of the researched 
exemplars made 50 pier. {\%}. Apparently, in process of propagation of 
single crystals concentration of impurity increases.

\begin{figure}
\includegraphics[width=1\linewidth]{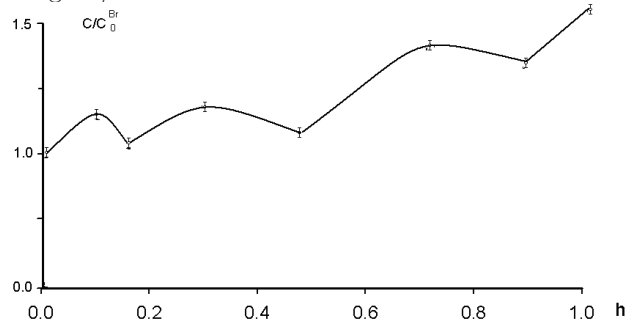}
\caption{Allocation of impurity of para-dichlorbenzol and p-BromoClorobenzene) longwise (h) the mixed chip at a high speed of propagation.}
\label{fig4}
\end {figure}

In Fig.~\ref{fig4} graphics of allocation of a builder of para-dichlorbenzol 
C/C$^{Br}_{0}$ on an axis of a single crystal in solid solutions are 
presented at greater growth rate V = 20.0 $\cdot $10-6 cm/s and a lapse rate 
of temperature dT/dl=7.7 grad/cm.

As we see Allocation of builders has wavy character. The velocity of pulling 
down of dish at which is broken monotone allocation makes in our case V 15.0 
$\cdot $10-6 cm/s. 

So, changing requirements of selection it is possible to change allocation 
of molecules of impurity thus probably as a uniform modification of 
concentration of all builders along an exemplar, and a wavy modification of 
concentration of two substances that will affect long-distance and 
short-range order of layout of molecules.

\end{document}